\documentclass[conference]{IEEEtran}
\IEEEoverridecommandlockouts
\usepackage{cite}
\usepackage{amsmath,amssymb,amsfonts}
\usepackage{algorithmic}
\usepackage{booktabs}
\usepackage{graphicx}
\usepackage{textcomp}
\usepackage{xcolor}
\usepackage{colortbl}
\usepackage{multirow}
\usepackage{array}
\usepackage{threeparttable}
\usepackage{subcaption}
\usepackage{url}
\usepackage{float}
\usepackage{balance}

\newcolumntype{C}{>{\centering}p{0.045\textwidth}}
\newcolumntype{X}{>{\centering\arraybackslash}m{0.045\textwidth}}

\def\BibTeX{{\rm B\kern-.05em{\sc i\kern-.025em b}\kern-.08em
    T\kern-.1667em\lower.7ex\hbox{E}\kern-.125emX}}
    
\usepackage[hyperfootnotes=false]{hyperref}    

\begin{document}

\title{Evaluation of Load Prediction Techniques for Distributed Stream Processing}

\author{
\IEEEauthorblockN{Kordian Gontarska\IEEEauthorrefmark{1}\IEEEauthorrefmark{2}, Morgan Geldenhuys\IEEEauthorrefmark{2}, Dominik Scheinert\IEEEauthorrefmark{2},\\ Philipp Wiesner\IEEEauthorrefmark{2}, Andreas Polze\IEEEauthorrefmark{1}, and Lauritz Thamsen\IEEEauthorrefmark{2}}
\IEEEauthorblockA{\IEEEauthorrefmark{1}Hasso Plattner Institute, University of Potsdam, Germany, \{firstname.lastname\}@hpi.de}
\IEEEauthorblockA{\IEEEauthorrefmark{2}Technische Universit{\"a}t Berlin, Germany, \{firstname.lastname\}@tu-berlin.de}
}

\maketitle
\thispagestyle{plain}
\pagestyle{plain}

\begin{abstract}
Distributed Stream Processing (DSP) systems enable processing large streams of continuous data to produce results in near to real time. They are an essential part of many data-intensive applications and analytics platforms.
The rate at which events arrive at DSP systems can vary considerably over time, which may be due to trends, cyclic, and seasonal patterns within the data streams. A priori knowledge of incoming workloads enables proactive approaches to resource management and optimization tasks such as dynamic scaling, live migration of resources, and the tuning of configuration parameters during run-times, thus leading to a potentially better Quality of Service.

In this paper we conduct a comprehensive evaluation of different load prediction techniques for DSP jobs. 
We identify three use-cases and formulate requirements for making load predictions specific to DSP jobs. Automatically optimized classical and Deep Learning methods are being evaluated on nine different datasets from typical DSP domains, i.e. the IoT, Web 2.0, and cluster monitoring. 
We compare model performance with respect to overall accuracy and training duration.
Our results show that the Deep Learning methods provide the most accurate load predictions for the majority of the evaluated datasets.
\end{abstract}

\begin{IEEEkeywords}
Distributed Stream Processing, Resource Management and Optimization, Load Prediction, Time Series Forecasting, Machine Learning
\end{IEEEkeywords}

\section{Introduction}
\label{sec:introduction}

Distributed Stream Processing (DSP) systems are responsible for extracting valuable insights from large streams of real-time data. 
Application areas include IoT data processing, click stream analysis, network monitoring, fraud detection, spam filtering, and news processing~\cite{IsahAMAZK19,NNG19}. 
These jobs require high throughput rates and low end-to-end processing latencies to support time-sensitive decisions. 
Input streams, however, are dynamic in nature and processing loads, therefore, have the potential to change significantly over time. 
Consequently, DSP systems like Heron~\cite{KBF+15}, Spark~\cite{ZahariaCFSS10} and Flink\cite{CarboneKEMHT15} are able to scale horizontally across a cluster of commodity nodes in order to accommodate different processing loads.
With this potential for workloads to fluctuate over time and the ability of systems to dynamically adjust configurations at runtime, research into approaches for the adaptive management of resources has grown in popularity in recent times.

Prominent applications within this domain include: 
the automated dynamic scaling of resources which aims to reduce over- and under-provisioning, minimizing operating costs and preventing possible reductions in the service quality \cite{GedikSHW14,FloratouAGRR17,Kalavri18,huKZ19};
The live migration of functionality/state across the network where cluster metrics are collected, processed, and scanned for anomalies which might reveal performance degradations and/or signs of component failure\cite{liu2014virtual,MasdariK20a, Chakrabarti2020};
and the automatic system tuning where dynamic runtime adjustment of system configurations are performed in order to improve overall system availability and reliability\cite{Geldenhuys2019EffectivelyTS,JHK20,Geldenhuys2020ChironOF}.
The majority of these approaches rely on coarse-grained metrics to reactively make remediation decisions. 
In some cases these metrics are also system specific and/or require a customized version of the system.
Interestingly, an increasing number of approaches use Time Series Forecasting (TSF), a more generalizable and proactive basis for making good decisions.

% TSF is a technique used for predicting future observations, e.g. load, traditionally in areas such as network traffic, and electrical systems\cite{SvigeljSA15,sivakoti2015load}.
Through forecasting what is to come we are better able to anticipate and plan for the future. 
While the use of TSF is not new in the context of DSP systems, in most cases it requires experts to pick the appropriate method and configure it to the situation at hand.
% Additionally, a multitude of methods exist with no consensus for which will perform best overall or for any specific DSP workload. 
Moreover, to the best of our knowledge, no comprehensive comparison has been conducted thus far where previous attempts at comparing these methods were inconclusive and present results obtained from a small number of datasets\cite{Kim16}.

In this paper we evaluate seven methods which meet the requirements for performing TSF for DSP jobs across nine different streaming datasets. 
We select various classical TSF methods as well as two Deep Learning (DL) methods. 
The datasets were sampled across various application domains, including: IoT, Web 2.0, and cluster monitoring. 
Each consists of a single value representing the number of events arriving at the DSP system every second and thus represents the load upon the system.
Based on the three main use cases: dynamic scaling, live migrations, and system tuning; we investigate three sampling rates: the first for more shorter term higher granularity predictions, the second for mid-range predictions; and the third for more longer term lower granularity predictions. 
Critically, we provide implementations for automatically optimizing the hyperparameters of the prediction models thus removing the necessity for configuring them by hand.
% Training and prediction durations were measured and are presented as part of the results.

% \pagebreak  % FIX

The remainder of the paper is structured as follows: Section II identifies the use cases and requirements for performing TSF for DSP systems. Section III presents the TSF methods, experimental design, evaluation metrics, time series datasets, and the experimental setup. Section IV presents and discusses our results. Section V describes the related work on TSF for load prediction, while Section VI concludes the paper.

\section{Problem Analysis}
\label{sec:problem_analysis}

In this section we first motivate the use of TSF for DSP systems by exploring different use cases and then explain the requirements for TSF for such tasks.

\subsection{Use Cases}
\label{sec:use_cases}

We identified three use cases where accurate load prediction would be quite useful for efficiently running DSP jobs despite varying input rates.

\subsubsection{Dynamic Scaling}

    The scalability of DSP systems is achieved by partitioning the entire data streams among data-parallel task instances that are deployed across clusters of computers.
    % However, this requires a decision on the number of partitions and the number of cluster nodes to be used for the processing of one input stream.
    % At the same time, streaming jobs are by nature long-running and streaming workloads often change over time. 
    Streaming jobs are by nature long-running and streaming workloads often change over time. 
    % Therefore, dynamic scaling of the parallelism and allocated resources may be required to maintain performance. 
    Many DSP systems have the ability to scale dynamically, yet the difficult task of deciding by how much and when is typically still left up to the user with off-the-shelf DSP systems. 
    Furthermore, the current research on this problem relies on coarse-grained metrics such as CPU utilization, throughput, and backpressure which tend to show incorrect provisioning, oscillations, and long convergence times\cite{Kalavri18}. 
    There is, therefore, a promising opportunity to combine TSF with performance modeling in order to scale parallelism and resource usage proactively based on predicted loads.
    For this use case a shorter prediction time horizon is important, i.e. 5 minutes.
    % i.e. if the system should scale up or down, then knowing what the expected average load would be at a 5 minute interval is of most value.
    
    \subsubsection{Timing of Live Migrations}
    
    As DSP systems approach even greater scales, the number of things that can go wrong increase as well.
    Live migrations allow administrators to transparently tune the performance of computing infrastructures when performance degradations in cluster nodes are detected, migrating system functionality and state over the network.
    Much of the research in this area is focused on the methods for detecting these anomalies, however, the question of when these preemptive actions should take place is likewise an important question to answer. 
    % In most cases this is left to the user to decide. 
    Here TSF can provide a way to automate this process by predicting periods of low utilization where changes to the cluster will have potentially less of an impact on overall performance.
    For this use case a mid-range prediction time horizon, i.e. 15 minutes, is important.
    
    \subsubsection{Automatic System Tuning}
    Self-tuning systems are those which are capable of optimizing their own internal configuration parameters in order to maximize or minimize the fulfilment of one or more objective functions. 
    If done correctly, this can yield systems which are more performant, cost effective, reliable, and power efficient.
    In order to achieve this, approaches within this domain commonly conduct profiling runs and/or use historical data in order to model performance behaviors.
    TSF, when combined with these models, not only allows the system to determine which configurations need tuning and by how much, but critically also the timings of these updates.
    As workloads change over time, delaying configuration updates can greatly improve the effectiveness of optimizations.
    For this use case, longer-term forecasts with hourly time horizons are most useful.
    
    %TODO: Some nice description of what that is, why it's important and obv why TSF may help in timing the optimization as it is not critical to the systems functionality/availability.
    %TODO: Sampling rate of 1 hour. Forecast horizon of 12 h. Find the hour with the lowest expected util to schedule a change in config params.

\subsection{Requirements}

DSP systems come with their own specific requirements which need to be taken into consideration while designing approaches to load prediction. In this section we define these requirements and present them as follows: 

    \subsubsection{Minimal Configuration}
    Solutions should follow a plug-and-play approach. User configuration requirements for tuning model hyperparameters should be kept to an absolute minimum. As input data streams and infrastructures deviate greatly, solutions should provide automatic optimization mechanisms to quickly adapt to a wide range of deployments.
    
    \subsubsection{Limited Model Inputs}
    Inputs consist of the ingress rate values, i.e. the sum of the events entering the source operators of the DSP job per second. As this is a scalar value, the focus should be on univariate time series prediction. Concerning the amount of data that is used for initial training, this depends on how long the DSP job has been running and availability of metrics data. 
    % DSP systems generally produce large volumes of metrics data and it is not always feasible to expect everything to be stored indefinitely. 
    The default retention periods of time series databases such as InfluxDB\footnote{\url{https://www.influxdata.com/}, accessed 2020-04-12} and Prometheus\footnote{\url{https://prometheus.io/}, accessed 2020-04-12} commonly used for storing such data is 7 and 15 days respectively. Thus, a limited amount of training data should be expected.
    
    \subsubsection{Frequent Model Updates}
    As new observations are continually being produced, models should consider these for their forecasts such that the quality of the predictions does not degrade. This can be achieved by periodically retraining the models or updating them before each new prediction with the latest observations produced since the last prediction was made. The parameters in classical methods need to be updated to the most recent observations to capture the changes in the time series. Opposed to that ML models inherently incorporate the most recent observations to perform inference, making updates to the model weights not as urgent.
    % Because of this, it is reasonable to assume that a trade-off between the computational overhead, energy consumption, and duration of retraining/updating the models needs to be found against model choice and their performance. Importantly, no resource guarantees can be made of what the underlying hardware upon which the solution is deployed should contain, i.e. a selected method should take advantage of a GPU if it is available, otherwise fallback to the CPU.
    
    \subsubsection{Variable Forecasting Length}
    Models which support multi-step time series forecasting are required. Understanding how expected workloads change over time is an important aspect to consider when designing strategies for the use cases which we have previously defined. Solutions should have the ability to incorporate various sample rates to match desired time horizons.
\section{Methodology}
\label{sec:experiment_methodology}

In this section we present our experiment methodology by introducing TSF methods we investigate, time series datasets upon which we base our comparison, the evaluation metrics by which we measure the performance of our models, and the setup of our experiments. The code and datasets for our experiments can be found in the associated github repository\footnote{Available at \url{https://github.com/dos-group/load-prediction-dsp}}.

\subsection{TSF Methods}
The following are methods commonly used in the domain of TSF. One of the targets of our investigation is to validate whether or not these methods are suitable considering the requirements defined and presented in the previous section. These methods are as follows:

    \subsubsection{Baselines - Last Observation and Last Day Observation}
    The two performance baseline models are established by taking either the last observation or the observation from before 24 hours, and projecting it forwards for the length of the forecast.
    
    \subsubsection{SES}
    Single Exponential Smoothing (SES) introduces exponentially decreasing weights, allowing to prioritize more recent observations and thus capture latest trends ~\cite{Holt2004ForecastingSA,winters1960forecasting}. 
    % We perform a gridsearch to find the optimal configuration.
    
    \subsubsection{TES}
    Holt and Winters~\cite{Holt2004ForecastingSA,winters1960forecasting} extended SES and developed Triple Exponential Smoothing (TES) which allows incorporating trend and seasonality. 
    % We find the optimal configuration via grid search. 
    
    \subsubsection{ARIMA}
    Auto Regressive Integrated Moving Average (ARIMA) is a widely used approach for time series forecasting~\cite{hyndman2018forecasting}.
    
    \subsubsection{SARIMA}
    Due to the seasonal nature of the datasets, we included Seasonal Auto Regressive Integrated Moving Average (SARIMA) models, which add seasonality to ARIMA models~\cite{hyndman2018forecasting}. 
    % The hyperparameters of the model are searched for by the same Auto-ARIMA step-wise optimizer.
    
    \subsubsection{Prophet}
    Prophet \cite{prophet} is a popular open source time series forecasting library developed by Facebook. It aims at making time-series data analysis available to non-expert users.
    
    \subsubsection{CNN}
    Convolutional Neural Network (CNN) architectures have been used successfully for time series prediction on multiple occasions~\cite{WangLZHCLC19,KoprinskaWW18}, as they allow for increased training speed and deep architectures due to shared filters. 

    \subsubsection{GRU}
    Recurrent Neural Networks (RNNs) allow for the memorization of past observations which makes them suitable for applications on time series~\cite{fu2016using,HuaZLCLZ19}. We design a gated recurrent unit (GRU)~\cite{ChungGCB14}.

\subsection{Evaluation Metrics}
Our main objective is the Symmetric Mean Absolute Percentage Error (SMAPE), which is commonly used for forecasting problems when dealing with data sources of varying value scales~\cite{shcherbakov2013survey, hyndman2006another, kreinovich2014estimate}. SMAPE is defined as

\begin{equation}
\label{eq:1}
\small
  SMAPE = \frac{100\%}{n} \sum_{t=1}^{n} \frac{|Y\textsubscript{t}-\hat{Y}\textsubscript{t}|}{|Y\textsubscript{t}|+|\hat{Y}\textsubscript{t}|}
\end{equation}

and delivers an error value between 0\% and 100\%. It is furthermore utilized as the loss function for our chosen DL methods as it is invariant to absolute errors. We also investigate another commonly used metric in regression tasks, the Root Mean Square Error (RMSE), which is defined as

\begin{equation}
\label{eq:2}
\small
  RMSE = \sqrt{\frac{1}{n}\sum_{t=1}^{n} (Y\textsubscript{t}-\hat{Y}\textsubscript{t})^2}
\end{equation}

 and allows us to assess the goodness of our predictions in terms of actual similarity to the true values.

\subsection{Time Series Datasets}

\begin{figure*}
    \centering
    \includegraphics[width=1\linewidth]{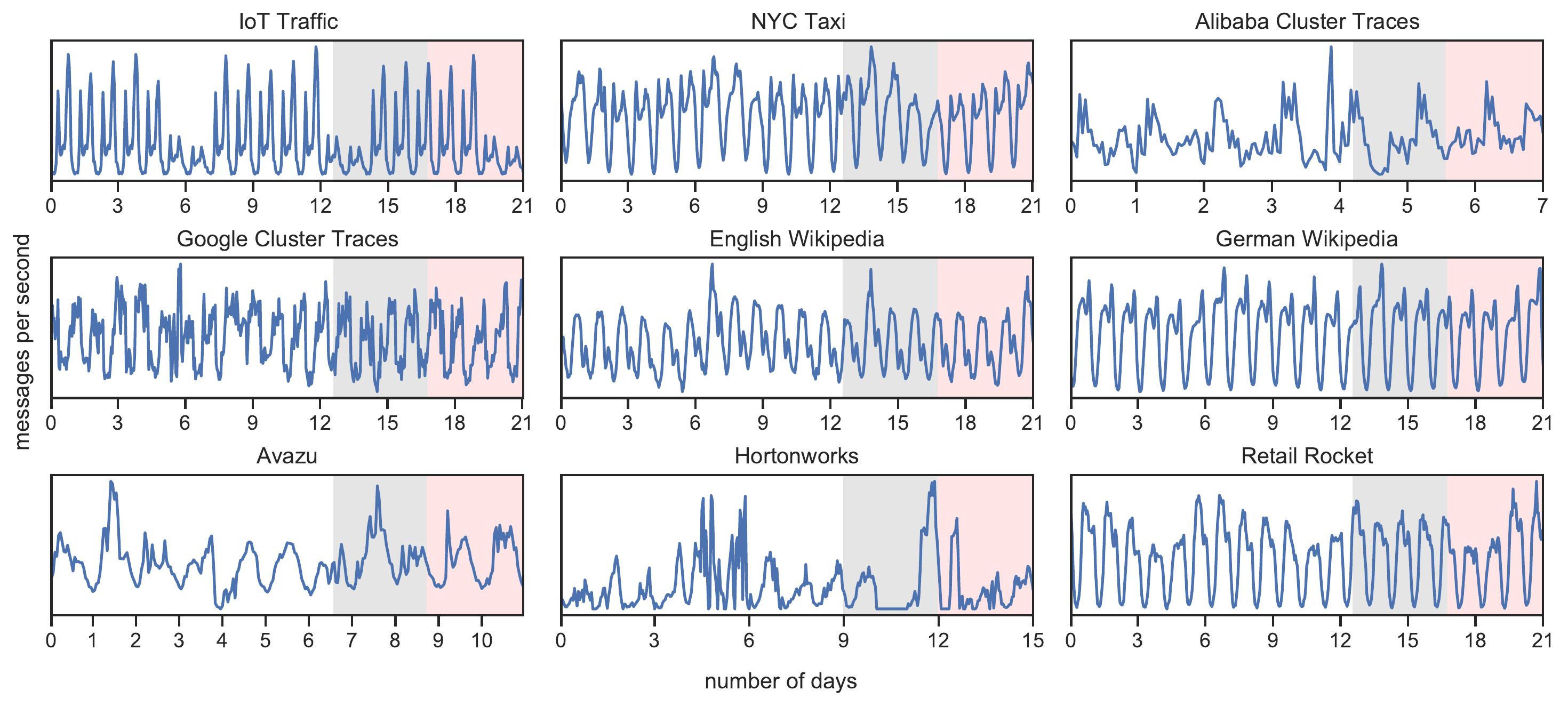}
	\caption{All datasets with hourly sampling rate highlighted as train (white), validation (gray), and test (red) data.}
    	\label{fig:datasets}
\end{figure*}

\label{sec:datasets}
We selected 9 heterogeneous and univariate datasets to investigate the applicability of our chosen methods on varying data. These datasets represent the expected load in a DSP system. The time series describe the sum of the events entering the source operators of a DSP job per second.

\subsubsection{IoT Traffic}
A synthetic dataset created using the SUMO~\cite{lopezBBEFHLRWW18} traffic simulator and the TAPAS cologne\footnote{\url{https://sumo.dlr.de/docs/Data/Scenarios/TAPASCologne.html}, accessed 2020-04-12} simulation scenario. The data represents the number of vehicles on the streets per second over time. 

\subsubsection{NYC Taxi} A dataset from the NYC Taxi \& Limousine Commission\footnote{\url{https://www1.nyc.gov/site/tlc/about/tlc-trip-record-data.page}, accessed 2020-04-12} on taxi rides in New York City. We aggregated the number of taxi rides of all providers during every five minutes over a three week period in January 2020.

\subsubsection{Alibaba Cluster Traces}
Production data originating from an Alibaba cluster\footnote{\url{https://github.com/alibaba/clusterdata}, accessed 2020-04-12} in 2018. We extract the start time of batch tasks over a 7-day period and aggregate them.

\subsubsection{Google Cluster Traces}
Workload traces recorded from eight different Google Borg cells during the month of May 2019\footnote{\url{https://github.com/google/cluster-data}, accessed 2020-04-12}. From this trace we extracted information about instance events over a randomly selected period of 21 days.

\subsubsection{Wikipedia Pageviews}
We used the Wikipedia API\footnote{\url{https://wikimedia.org/api/rest\_v1}, accessed 2020-04-12} to extract the number of page views per hour for a period of three weeks in July 2019. Subsequently, we interpolate between the values to derive different sampling rates. The procedure is conducted both for the German and English Wikipedia, and thus results in two different datasets. 

\subsubsection{Avazu}
This dataset is created by using a click-through rate prediction dataset from Kaggle\footnote{\url{https://www.kaggle.com/c/avazu-ctr-prediction}, accessed 2020-04-12}, aggregating the clicks per hour over time, and linearly interpolating between the aggregated values to obtain different sampling rates.

\subsubsection{Hortonworks}
Clickstream events taken from a tutorial on visualizing clickstream data provided by Hortonworks\footnote{\url{https://www.cloudera.com/tutorials/visualize-website-clickstream-data.html}, accessed 2020-04-12}. We extracted the event timestamps on a per second granularity and aggregate them to the respective granularities.

\subsubsection{Retail Rocket}
E-commerce data containing events of the Retail Rocket recommender system\footnote{\url{https://www.kaggle.com/retailrocket/ecommerce-dataset}, accessed 2020-04-12}. We construct our dataset by aggregating behaviour data, such as click events, over a three week period.

\subsection{Dataset Adjustements}
All datasets have been adjusted to resemble realistic event rates in DSP jobs. Following from the use cases described in Section~\ref{sec:use_cases}, we resampled the datasets to sampling rates of 5 minutes, 15 minutes, and 1 hour.

We assume normally distributed noise in the datasets. In order to estimate the expected relative error from the datasets with original granularity of only 1 hour, we made use of the aforementioned datasets with 5-minute granularity. First, we used their resampled variant with sampling rate of 1 hour and resampled it to a 5-minute sampling rate. Next, we interpolated the missing observations using a second order spline function and calculated the standard deviation of the differences between the interpolated 5-minute sampling rate dataset and the original 5-minute granularity dataset. The expected relative error is $\sigma_e \approx 1 \%$.

Subsequently, we selected the datasets with original granularity of 1 hour and resampled them to a sampling rate of 5 minutes. The missing observations were interpolated. To simulate the aforementioned normally distributed noise we sampled $i$ relative errors from a normal distribution $\mathcal{N}\sim(0,\sigma_e)$, where $i$ is the amount of observations in the respective dataset. These sampled errors were then multiplied with the corresponding observations and added to them. Finally, we resampled these new datasets also to a sampling rate of 15 minutes.
    
Additionally, we scaled and shifted all datasets to have a mean message/second rate of $\mu \approx 108\,000$, and a standard deviation of $\sigma \approx 12\,000$. These lie mostly within a range of approximately $90\,000$ -- $150\,000$. 
This was done to make absolute metrics comparable across the datasets.
These adjustments result in the final datasets at three different sampling rates. 
% The impact of the different sampling rates is shown in Figure~\autoref{fig:sampling_rate}. 

% \begin{figure}
%     \centering
%     \includegraphics[width=1\linewidth]{images/sampling_rate.pdf}
% 	\caption{Two days of Google cluster traces data in 1 hour, 15 minutes, and 5 minutes sampling rate.}
%     	\label{fig:sampling_rate}
% \end{figure}

\subsection{Experimental Setup}
The training of the models as well as inference was conducted on a dedicated machine equipped with a GPU. Specifications and software versions can be found in~\autoref{clusterspecs}. 

\begin{table}[ht]
\centering
\caption{Specifications}
    \begin{tabular}[t]{rp{0.68\linewidth}}
        \toprule
        Resource&Details\\
        \midrule
        CPU, vCores & Intel(R) Xeon(R) Silver 4208 CPU @ 2.10GHz, 8\\
        Memory & 45 GB RAM\\
        GPU & 1 x NVIDIA Quadro RTX 5000 (16 GB memory)\\
        Software & PyTorch 1.7.0, PyTorch Ignite 0.4.2\\ 
        & Ray Tune 1.1.0, Optuna 2.3.0\\
        & pmdarima 1.8.0, statsmodels 0.12.2, fbprophet 0.7.1\\
        \bottomrule
    \end{tabular}
\label{clusterspecs}
\end{table}%

\subsubsection{Experiment Execution}
Each of our experiments examines the forecasting capabilities of a certain method on a specific dataset. 
We conduct 189 experiments overall, exploring seven methods on nine datasets with three different sampling rates.
For each sampling rate, we define a different forecast horizon: For the 5 minute interval, we aim at predicting only the next value, the 15 minute interval comes with a horizon of four values, and the one hour interval is associated with a prediction horizon of twelve. This reflects our different use cases and the corresponding prediction capabilities. 

For the DL methods, i.e. CNN and GRU, we split each dataset along the time domain into training, validation, and test subset (60\%, 20\%, 20\%). The validation subset is used during hyperparameter optimization to abort unpromising trials and to determine the best trial. All other methods conduct a split into training and test subset (80\%, 20\%) and thus effectively train on more data. This is illustrated in~\autoref{fig:datasets}.

\subsubsection{Training and Model Configuration}
%In the following, we describe how the respective models are configured, what the hyperparameter searchspaces look like and how the training is conducted.

We fit the level smoothing factor $\alpha$ of SES models based on their achieved Akaike information criterion (AIC).
For TES models, we additionally fit $\beta$ and $\gamma$, the smoothing factors for trend and seasonality. Furthermore, we examine the effect of assuming an additive or multiplicative seasonality.

For training ARIMA models, we perform a hyperparameter search on $p$ and $q$, and test the order of first-differencing with $d = [0,1]$. For fitting SARIMA models, the additional seasonal hyperparameters $P$, $D$, and $Q$ are integrated into the hyperparameter search. We set boundaries for $P$ and $Q$, and additionally search explicitly the space defined by $d \times D = [0,1] \times [0,1]$. In the process, we assume a seasonal period of 24 hours. For both ARIMA and SARIMA, the AIC estimator is used to determine the best hyperparameters.

For Prophet, the hyperparameter optimization and fitting is handled entirely by the library. 
% Additionally we minimized the forecasting time by setting the prediction of confidence intervals to false as we are only interested in the mean forecast.

For the machine learning models we search for optimal configurations of a variety of different hyperparameters such as number of layers, number of neurons, learning rate, and dropout rate. 
For both CNN and GRU, we are moreover especially interested in the optimal length of input sequences extracted from the training data. 
The details of the spanned hyperparameter searchspaces are described in our repository, along with further technical details.
\section{Experiment Results}
\label{sec:experiment_results}

\begin{figure*}
    \centering
    \begin{subfigure}{.46\textwidth}
      \centering
      \includegraphics[width=1\linewidth]{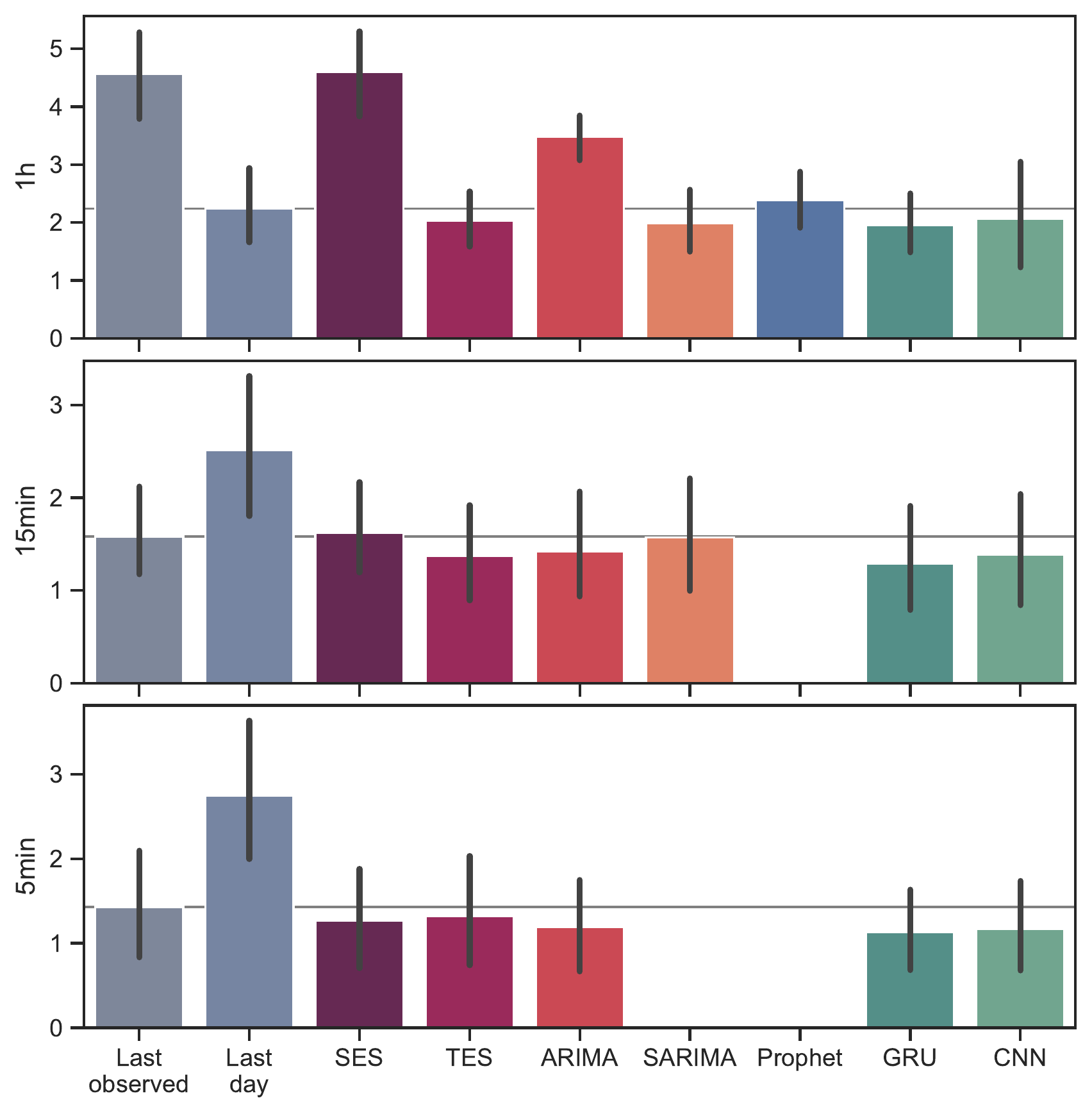}
	    \caption{Mean SMAPE results across all datasets.}
      \label{fig:results_smape}
    \end{subfigure}
    \begin{subfigure}{.48\textwidth}
      \centering
      \includegraphics[width=1\linewidth]{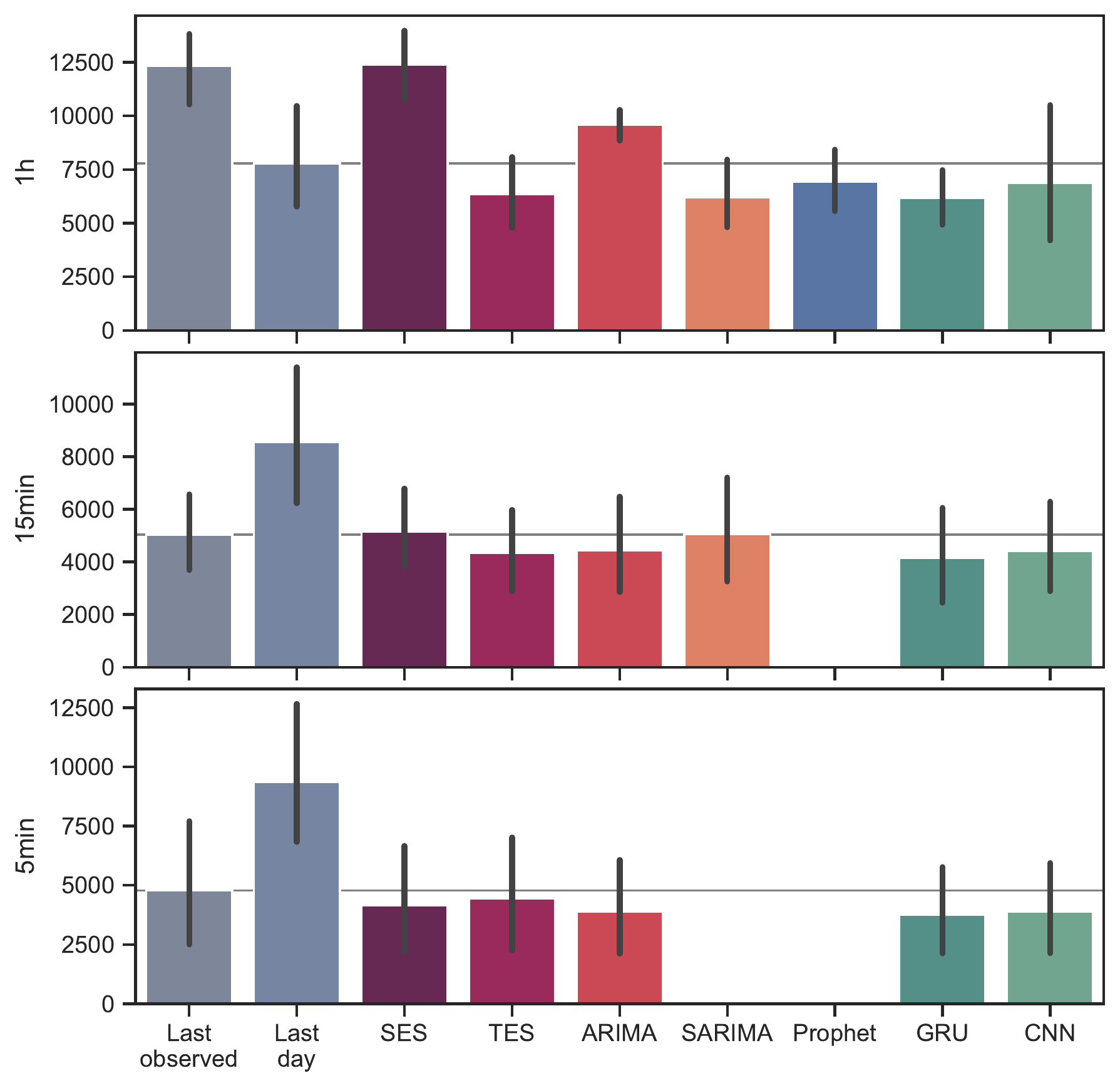}
	    \caption{Mean RMSE results across all datasets.}
      \label{fig:results_rmse}
    \end{subfigure}
    \caption{Performance of methods evaluated by the SMAPE and RMSE metrics.}
    \label{fig:results}
\end{figure*}

\begin{center}
\begin{table*}
\caption{SMAPE of all models on the different datasets and sampling rates.}
\centering
\begin{threeparttable}
\begin{tabular}{c||l|CC|CCCCCCX}
\toprule
     &  & Last &  Last  &   &   &   &   &   &   &  \\
     &  &  obs. &  day & SES &  TES &  ARIMA &  SARIMA &  Prophet &  GRU &  CNN   \\
\midrule
\parbox[t]{2mm}{\multirow{9}{*}{\rotatebox[origin=c]{90}{\textbf{1 hour}}}} 
     & IoT Traffic &           4.21 &      1.44 &        4.21 &         2.21 &   3.53 &    1.81 &     2.76 & 1.75 & \cellcolor{blue!20}1.02 \\
     & NYC Taxi &           5.18 &      1.46 &        5.18 &         \cellcolor{gray!20}1.27 &   3.65 &    \cellcolor{gray!20}1.15 &     2.49 & \cellcolor{gray!20}1.21 & \cellcolor{blue!20}0.83 \\
     & Alibaba &           2.66 &      2.29 &        2.69 &         \cellcolor{gray!20}2.23 &   \cellcolor{gray!20}2.24 &    \cellcolor{blue!20}1.88 &     2.46 & 3.08 & 3.21 \\
     & Google &           5.44 &      3.19 &        5.56 &         \cellcolor{gray!20}2.65 &   4.46 &    \cellcolor{gray!20}2.61 &     \cellcolor{gray!20}2.89 & \cellcolor{blue!20}2.46 & \cellcolor{gray!20}2.89 \\
     & Wikipedia EN &           4.86 &      1.39 &        4.86 &         \cellcolor{blue!20}1.10 &   3.36 &    \cellcolor{gray!20}1.18 &     \cellcolor{gray!20}1.34 & \cellcolor{blue!20}1.10 & \cellcolor{gray!20}1.33 \\
     & Wikipedia DE &           6.14 &      1.28 &        6.14 &         1.39 &   3.97 &    \cellcolor{gray!20}1.23 &     1.39 & \cellcolor{gray!20}1.11 & \cellcolor{blue!20}0.65 \\
     & Avazu &           3.53 &      2.61 &        3.53 &         \cellcolor{gray!20}2.35 &   3.53 &    \cellcolor{gray!20}2.16 &     \cellcolor{gray!20}1.82 & \cellcolor{blue!20}1.77 & \cellcolor{gray!20}1.87 \\
     & Hortonworks &           3.25 &      4.43 &        3.41 &         3.51 &   \cellcolor{blue!20}2.94 &    3.84 &     3.84 & 3.38 & 5.27 \\
     & Retail Rocket &           5.79 &      2.07 &        5.80 &         \cellcolor{gray!20}1.64 &   3.63 &    \cellcolor{gray!20}2.06 &     2.46 & \cellcolor{gray!20}1.72 & \cellcolor{blue!20}1.51 \\
\midrule
\parbox[t]{2mm}{\multirow{9}{*}{\rotatebox[origin=c]{90}{\textbf{15 minutes}}}} 
     & IoT Traffic &           1.27 &      1.46 &        1.27 &         \cellcolor{gray!20}1.14 &   \cellcolor{gray!20}0.83 &    \cellcolor{gray!20}0.70 & - & \cellcolor{blue!20}0.47 & \cellcolor{gray!20}0.54 \\
     & NYC Taxi &           1.16 &      1.40 &        1.17 &         \cellcolor{gray!20}0.49 &   \cellcolor{gray!20}0.74 &    \cellcolor{blue!20}0.40 & - & \cellcolor{gray!20}0.44 & \cellcolor{gray!20}0.52 \\
     & Alibaba &           2.81 &      2.92 &        2.85 &         \cellcolor{gray!20}2.42 &   \cellcolor{gray!20}2.63 &    \cellcolor{blue!20}2.21 & -  & \cellcolor{gray!20}2.72 & \cellcolor{gray!20}2.69 \\
     & Google &           \cellcolor{blue!20}3.01 &      4.30 &        3.19 &         3.06 &   3.27 &    3.46 & - & 3.02 & 3.31 \\
     & Wikipedia EN &           1.08 &      1.71 &        1.11 &         \cellcolor{blue!20}0.70 &   \cellcolor{gray!20}0.90 &    \cellcolor{gray!20}0.91 & - & \cellcolor{gray!20}0.75 & \cellcolor{gray!20}0.78 \\
     & Wikipedia DE &           1.31 &      1.48 &        1.31 &         \cellcolor{gray!20}0.69 &   \cellcolor{gray!20}0.85 &    \cellcolor{gray!20}0.78 & - & \cellcolor{blue!20}0.63 & \cellcolor{gray!20}0.83 \\
     & Avazu &           1.11 &      2.53 &        1.11 &         1.35 &   \cellcolor{blue!20}1.02 &    1.49 & - & 1.13 & 1.17 \\
     & Hortonworks &           \cellcolor{blue!20}1.09 &      4.48 &        1.11 &         1.36 &   1.18 &    1.83 & - & 1.38 & 1.44 \\
     & Retail Rocket &           1.39 &      2.30 &        1.49 &         \cellcolor{gray!20}1.12 &   1.39 &    2.34 & - & \cellcolor{blue!20}1.09 & \cellcolor{gray!20}1.21 \\
\midrule
\parbox[t]{2mm}{\multirow{9}{*}{\rotatebox[origin=c]{90}{\textbf{5 minutes}}}} 
     & IoT Traffic &           0.21 &      1.46 &        0.21 &         \cellcolor{gray!20}0.15 &   \cellcolor{blue!20}0.07 &  - & - & \cellcolor{gray!20}0.13 & \cellcolor{gray!20}0.10 \\
     & NYC Taxi &           0.34 &      1.40 &        0.34 &         \cellcolor{blue!20}0.26 &   \cellcolor{gray!20}0.31 & - & - & \cellcolor{gray!20}0.28 & \cellcolor{gray!20}0.28 \\
     & Alibaba &           3.47 &      3.30 &        \cellcolor{gray!20}2.92 &         \cellcolor{gray!20}3.19 &   \cellcolor{gray!20}2.53 & - & - & \cellcolor{blue!20}2.16 & \cellcolor{gray!20}2.37 \\
     & Google &           2.91 &      4.96 &        \cellcolor{gray!20}2.82 &         2.96 &   \cellcolor{gray!20}2.77 & - & - & \cellcolor{blue!20}2.56 & \cellcolor{gray!20}2.69 \\
     & Wikipedia EN &           1.19 &      1.96 &        \cellcolor{gray!20}0.97 &         \cellcolor{blue!20}0.93 &   \cellcolor{gray!20}0.94 & - & - & \cellcolor{blue!20}0.93 & \cellcolor{gray!20}0.96 \\
     & Wikipedia DE &           1.10 &      1.73 &        \cellcolor{gray!20}0.96 &         \cellcolor{gray!20}1.00 &   \cellcolor{gray!20}0.90 & - & - & \cellcolor{gray!20}0.89 & \cellcolor{blue!20}0.87 \\
     & Avazu &           1.12 &      2.71 &        \cellcolor{gray!20}0.99 &         \cellcolor{gray!20}1.02 &   \cellcolor{blue!20}0.98 &  - & - & \cellcolor{gray!20}1.00 & \cellcolor{gray!20}1.02 \\
     & Hortonworks &           0.90 &      4.54 &        \cellcolor{blue!20}0.80 &         0.93 &   \cellcolor{gray!20}0.82 & - & - & \cellcolor{gray!20}0.86 & \cellcolor{gray!20}0.87 \\
     & Retail Rocket &           1.65 &      2.70 &        \cellcolor{gray!20}1.39 &         \cellcolor{gray!20}1.45 &   \cellcolor{gray!20}1.40 & - & - & \cellcolor{blue!20}1.38 & \cellcolor{blue!20}1.38 \\
\bottomrule
\end{tabular}
\begin{tablenotes}
 \item For each line, the best performing result is highlighted in blue. Results highlighted in grey beat both baseline models.
\end{tablenotes}
\end{threeparttable}
\label{tbl:performance_results}
\end{table*}
\end{center}

\subsection{Limitations}
Since Prophet is primarily designed to detect long term seasonality, it appears to struggle with short sampling rates. When fitting it to data with 5 minute or 15 minute sampling rate, the library only predicts hourly data and interpolates between these data points during prediction. We thus decided to only include the hourly sampling rate model into our evaluation.

The coefficients for the SARIMA models are being computed analytically. At a sampling rate of 5 minutes and a daily seasonality, one seasonal period consists of 288 observations, which renders the computation infeasible in our scenario. Hence, we excluded the evaluation of SARIMA on the 5 minute sampling rate datasets.

\subsection{Performance}

The condensed results plots in~\autoref{fig:results} together with the detailed insights given in~\autoref{tbl:performance_results} allow for a comprehensive discussion of the model performances.

From~\autoref{fig:results_smape}, we can infer that across all datasets and sampling rates, the DL methods are the best or among the best performing methods with a stable prediction performance. The stability aspect also holds true for TES, whereas all other classical methods are either experiencing difficulties for certain sampling rates, or are even not applicable to them.
The reported SMAPE results are also confirmed by the reported RMSE results in~\autoref{fig:results_rmse}. We have ensured to make the RMSE a valid metric by adjusting the datasets to be in the same range with a similar mean and standard deviation.

Summarizing~\autoref{fig:results}, we identify that the DL methods performed best across all use cases, with no universal alternative from the classical methods. We note that on average, TES outperforms our installed baselines across all sampling rates, while being comparably fast to train and stable in prediction performance. Meanwhile, ARIMA shows good prediction capabilities on the 5 minute and 15 minute sampling rate datasets, yet fails to capture seasonal patterns present in datasets with hourly sampling rate. 

Complementing the averaged scores in~\autoref{fig:results}, we present detailed results with~\autoref{tbl:performance_results}. They indicate that the DL methods exhibit a similarly good performance over all experiments. In the rare cases of inferior performance to our baselines, the performance degradation affects both models in the majority of situations, leading to the assumption that patterns found in the test data were not present in the training data. In fact, for two datasets, namely the Google and Hortonworks datasets with a sampling rate of 15 minutes, there is not a single model that is able to outperform the baseline, confirming our previous interpretation of certain results.

\begin{figure}
    \centering
    \includegraphics[width=1\linewidth]{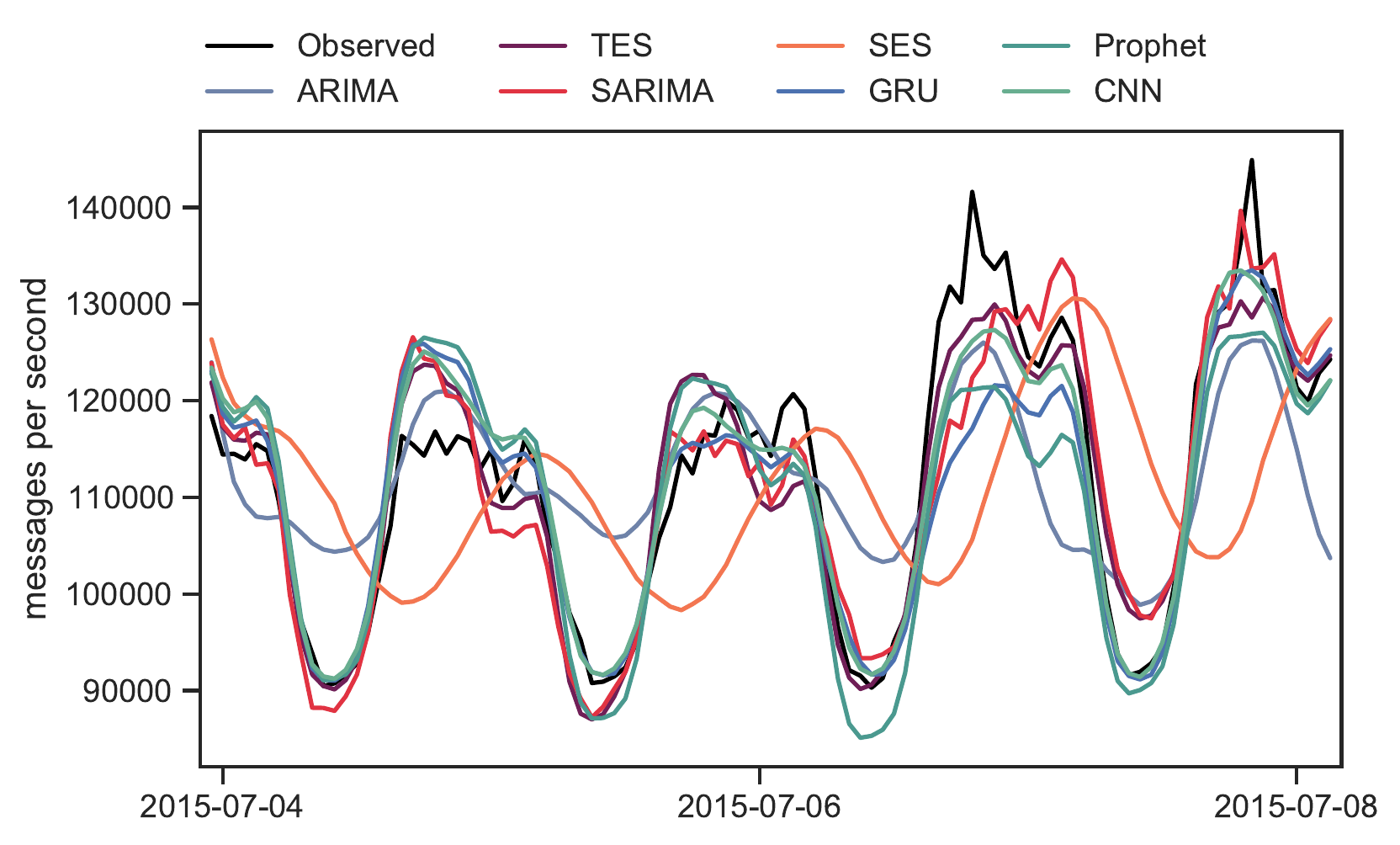}
	\caption{All model forecasts for the hourly sampled test data from Retail Rocket. Seasonality is captured to varying degrees.}
    	\label{fig:example}
\end{figure}

On the other hand, for 5-minute sampling rate, several models can be identified, each of which outperforms the baselines across all corresponding datasets. 
\autoref{fig:example} shows an example of the different model behavior for the hourly sampled test data from Retail Rocket.
It can be seen that the DL models as well as Prophet fit the seasonal pattern well, while the remaining classical models are more responsive to noise. 
Additionally, we note that methods which do not capture seasonality in their models perform poorly on the task to forecast the approaching 12 hours given hourly sampled data.
Interestingly, Prophet performed rather poorly in this task as well, failing to beat the baseline models in most cases.

The results further reveal that for hourly sampling and forecast horizon of 12 observations, CNN is among the best predictors with the exception of two outliers. 
SARIMA works well on hourly data as it has short seasonal patterns and thus states a good alternative from the classical methods, but struggles with long-term seasonality as present in the data sampled at 15 minutes. GRU are the best performing models for the use case of 15 minute sampling and a forecast horizon of four samples, and are closely followed by the CNNs. ARIMA and TES perform reasonably similar and make for good alternatives from the classical methods. While the GRU models perform best on predicting observations with a forecast horizon of 1 at a sampling rate of 5 minutes, ARIMA performs almost as well across all datasets. 

In conclusion, we find the DL methods as well as TES to be the most universal predictors, achieving good prediction results on the various datasets, sampling rates and hereby dataset sizes.

\subsection{Duration}

\begin{figure}
    \centering
    \includegraphics[width=0.95\linewidth]{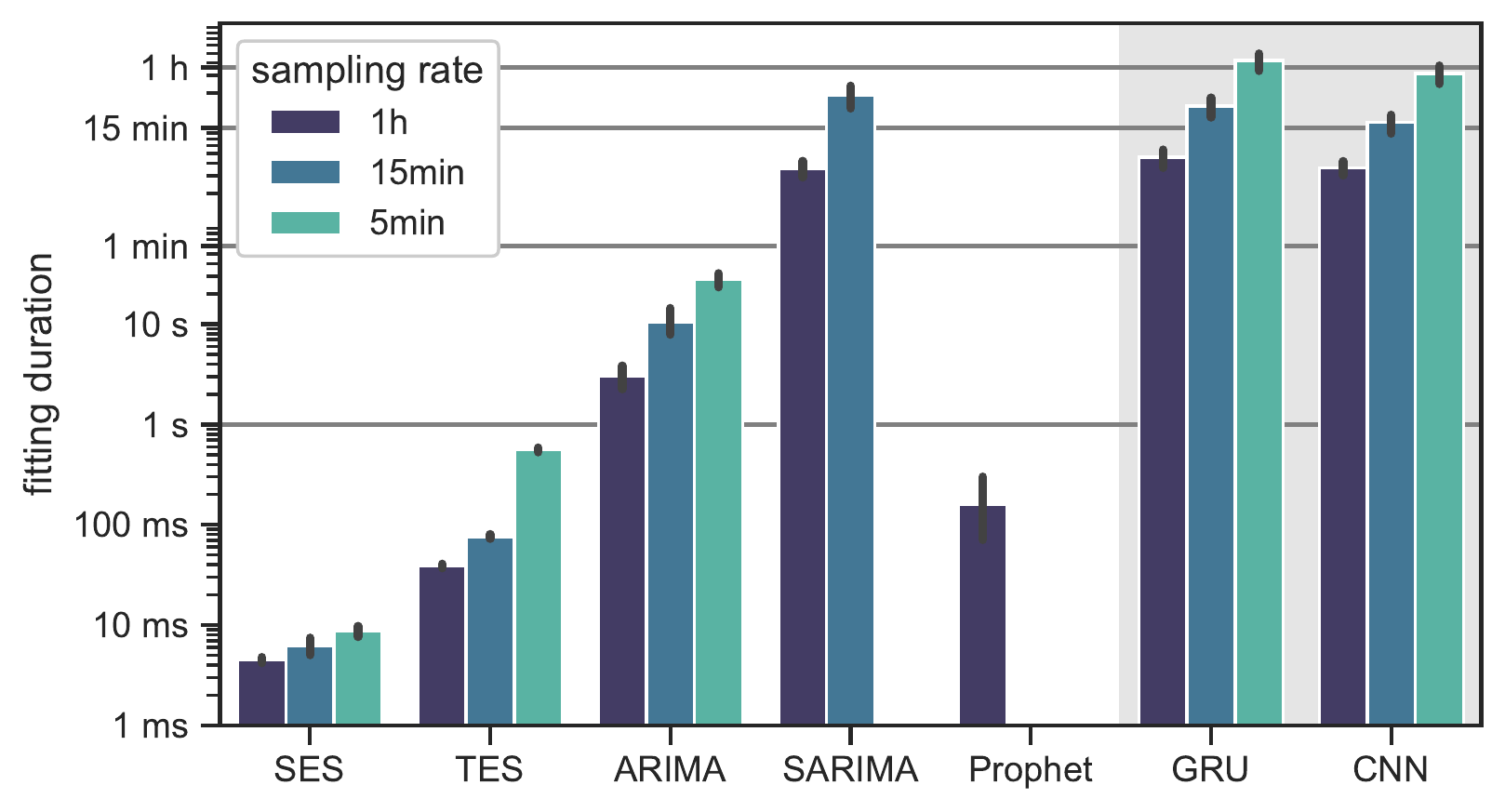}
    \includegraphics[width=0.95\linewidth]{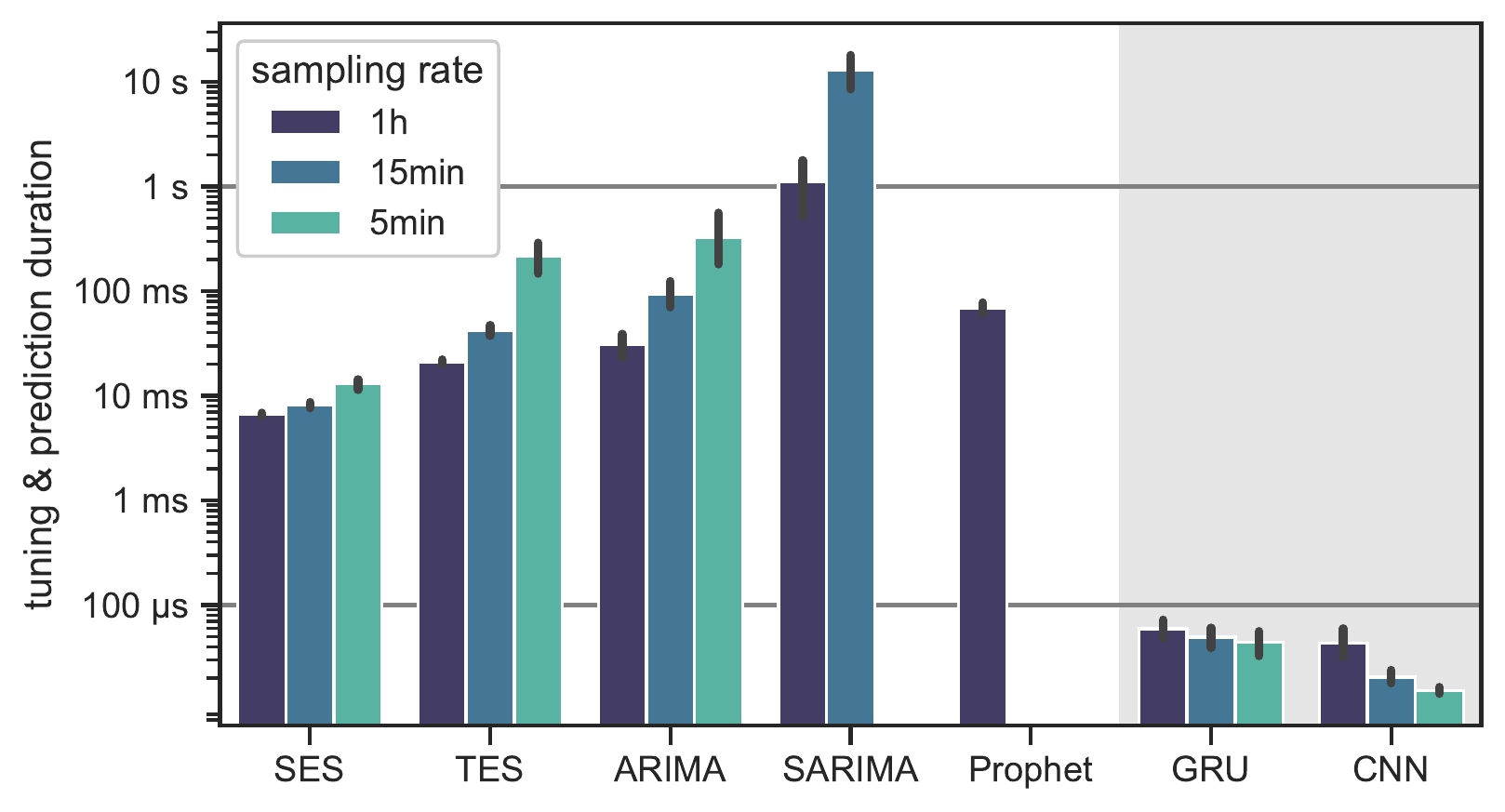}
	\caption{Duration of fitting the methods and performing predictions. Classical and DL methods are not directly comparable due to the different hardware used, hence the different background coloration.}
    	\label{fig:duration}
\end{figure}

We measured the time needed to find optimal hyperparameters and fit the respective models, as well as the average time it takes a model to predict the next observations and possibly fine-tune it to the most recent observation. The results are visualized in~\autoref{fig:duration}. As expected, the required time for training and fine-tuning depends on the amount of available training data, which is shown by the differences between the various sampling rates. 

It can be observed that there are significant differences in the time needed for initial fitting of the models. While the methods based on exponential smoothing are superior to the ARIMA variants, it can also be seen that the more complex variants of the respective methods (i.e. TES and SARIMA) require more time due to the extended hyperparameter search space. In comparison, Prophet is slightly more time-demanding than exponential smoothing, yet notably faster than the ARIMA methods. Although the DL methods are not directly comparable due to the utilization of different hardware, it is worth-mentioning that the training in general takes relatively long. Here, the training duration depends not only on the hyperparameter searchspace alone, but also on the specified configuration in terms of parallelism, maximum execution time, and preemptive termination of trials, as well as the total number of trials to investigate.

When using the trained models for prediction, it can be observed that all classical methods are significantly slower than the DL methods. The main reason for this is that the classical models need to be updated to the most recent observations to adapt to changes in the timeseries. This also explains the differences in time needed for the individual sample rates. This is in contrast to our DL methods, where the model parameters after the initial training are preserved and used for prediction, and the models only need to be supplied the most recent observations as input data for the context update.

In summary, the DL methods require a notable amount of time for initial training but are fast when doing prediction, whereas the classical methods are faster to fit while requiring time-consuming updates constantly over time.
\section{Related Work}
\label{sec:related_work}

TSF for load prediction is used by energy providers and electricity grid operators to plan ahead and meet demand and supply~\cite{ElectricityLoadForecastingReview_2020}. 
Nevertheless, it has also long been used in areas such as communication networks and cloud infrastructures. 
Understanding future load requirements for communication networks gives providers the ability to optimize resources allowing for better quality of service~\cite{ahmed2010empirical, cortez2012, jiangDSN19}. In cloud services, load prediction is used to balance resources within a data center as well as for scheduling live migrations~\cite{liu2014virtual, zhongZSG18, MasdariK20a}. 
Throughout all of these works, classical and DL methods have been used repeatedly. However, none has been directly compared under our defined requirements for performing TSF in DSP systems, such as minimal configuration and limited model inputs. In the context of DSP, TSF methods have been used in diverse forms and for varying reasons~\cite{muJLZW19,kalimCWLWLFQLCW19,wu2017fas,huKZ19,matteisM17}. While previous works successfully apply a selected method to a concrete problem, to the best of our knowledge, there is no related work that compares multiple TSF methods for DSP. Additionally, this paper formulates requirements for TSF for DSP and performs an evaluation on multiple real-world and one synthetic datasets from different domains.

\section{Conclusion}
\label{sec:conclusion}

This paper presented an evaluation of TSF methods for DSP systems. The methods we assessed include multiple classical and two DL models: SES, TES, ARIMA, SARIMA, Prophet, GRU, and CNN. The methods were evaluated across nine datasets from the IoT, Web 2.0, and cluster monitoring domains.
The results show that both CNN and GRU had the best prediction performance across the majority of the datasets, while requiring more time for initial model training. However, depending on the use case we found that there is always at least one classical model performing similarly well as the DL models, while being relatively fast to train. 
In general, the best performing models demonstrate to cope well even though the amount of training data was limited. In terms of configuration, it is more straightforward to optimally configure the diverse classical methods.

Based on our results, we deem both TES and the DL methods, and even a combination of these classes of approaches promising for use in the domain of DSP. Depending on the available resources and the use case at hand, one method might be preferable over the others.

\section*{Acknowledgments}
This work has been supported through grants by the German Federal Ministry for Economic Affairs and Energy (BMWi) as Telemed5000 (funding mark 01MD19014C), and by the German Federal Ministry of Education and Research (BMBF) as BIFOLD (funding mark 01IS18025A) and WaterGridSense 4.0 (funding mark 02WIK1475D).

\bibliographystyle{IEEEtran}
\balance
\bibliography{bib}

\end{document}